\documentclass{webofc}

\usepackage[varg]{txfonts}   
\usepackage{hyperref}
\usepackage{url}
\usepackage{comment}

\usepackage{mathrsfs}
\usepackage{amssymb}
\usepackage{amsmath}
\usepackage{graphicx}
\usepackage{amsfonts,amsbsy}
\usepackage{nicefrac}
\usepackage{xcolor}

\usepackage{setspace}   

\usepackage{bbm}

\usepackage{slashed}

\definecolor{lcolor}{rgb}{0.5,0,0}
\definecolor{citcolor}{rgb}{0,0.3,0.0}
\definecolor{ao(english)}{rgb}{0.0, 0.5, 0.0}

\definecolor{applegreen}{rgb}{0.55, 0.71, 0.0}
\definecolor{cadetblue}{rgb}{0.37, 0.62, 0.63}
\definecolor{cadet}{rgb}{0.33, 0.41, 0.47}
\definecolor{byzantine}{rgb}{0.74, 0.2, 0.64}
\definecolor{orange}{rgb}{1.0, 0.5, 0.0}

\def\bea{\begin{eqnarray*}}
\def\eea{\end{eqnarray*}}

\def\be{\begin{equation*}}
\def\ee{\end{equation*}}

\newcommand{\ud}{\mathrm{d}}

\hypersetup{colorlinks=true,citecolor=blue,urlcolor=blue,linkcolor=blue}
\begin{document}
\title{Overview of the latest developments in understanding the initial state and thermalization}

\author{\firstname{Kirill} \lastname{Boguslavski}\inst{1,2}\fnsep\thanks{\email{kirill.boguslavski@subatech.in2p3.fr}}
\fnsep\thanks{Invited plenary talk at the 31st International Conference on Ultra-relativistic Nucleus-Nucleus Collisions (Quark Matter 2025), Frankfurt, Germany, Apr 6-12, 2025.
\newline~
\newline {\em Acknowledgements:} I would like to thank all the speakers and colleagues on the topic of initial stages at Quark Matter 2025, especially with whom I had the pleasure to discuss their work. 
This work is funded in part by the Austrian Science Fund (FWF) under Grant DOI 10.55776/P34455.
}
}

\institute{SUBATECH UMR 6457 (IMT Atlantique, Université de Nantes, IN2P3/CNRS), 
\newline 4 rue Alfred Kastler, 44307 Nantes, France
\and
Institute for Theoretical Physics, TU Wien, Wiedner Hauptstraße 8-10, 1040 Vienna,
Austria
          }

\abstract{
A proper description of the non-equilibrium matter preceding the quark-gluon plasma (QGP) in heavy-ion collisions and its observable consequences remain a major theoretical challenge, while at the same time offering new opportunities for experimental exploration.
In these proceedings, I provide an overview of studies presented in talks and posters at Quark Matter 2025 on this topic. We will focus on the latest developments regarding the features and the numerical description of the non-equilibrium pre-QGP matter, as well as the potential to use hard probes as a means to study the hydrodynamization dynamics of the QCD plasma.
}
\maketitle
\section{Introduction}
\label{sec:intro}

In high-energy heavy-ion collisions, strongly interacting matter formed by quarks and gluons is created under extreme conditions between the collided nuclei, and it is described by Quantum Chromodynamics (QCD). 
As the medium cools during its evolution, it undergoes different phases, starting from an early pre-equilibrium regime, followed by a fluid-like quark-gluon plasma (QGP) phase, and eventually hadronization. The pre-equilibrium or pre-QGP phase is of particular interest since it probes the dynamical and quantum aspects of QCD.

\begin{figure}[t]
\centering
\includegraphics[width=0.9\textwidth]{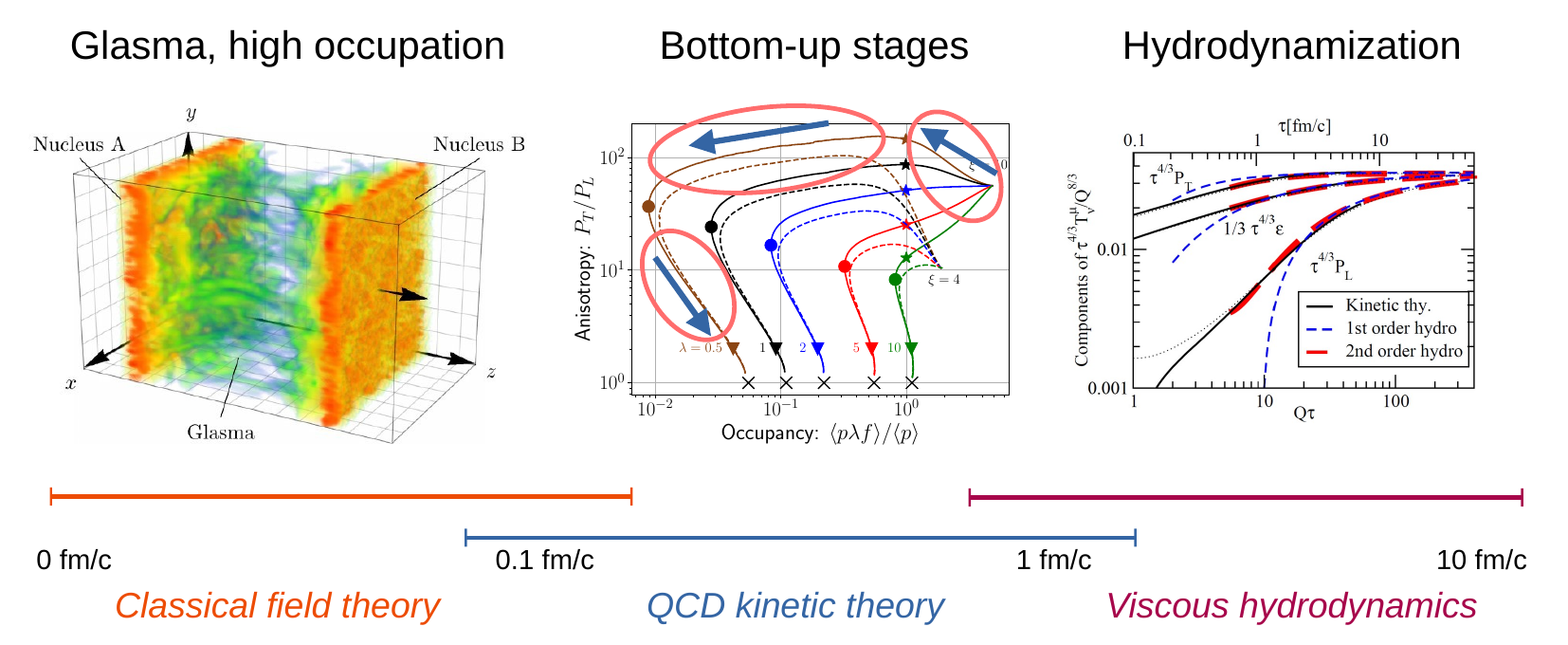}
\caption{An overview of different stages of the pre-QGP medium from the weak-coupling perspective including their physical interpretation and theoretical description. (figures from \cite{Ipp:2017lho, Kurkela:2015qoa})}
\label{fig:IS_QM}       
\end{figure}

From this perspective, the overarching goal is to learn about the real-time properties of QCD in such extreme environments. Two central research questions arise. First, what are the details of the initial stages of the pre-QGP in heavy-ion collisions? Considerable progress has been made in recent years through the interplay of initial-state modeling, classical-statistical simulations, and QCD kinetic theory, which together form the standard weak-coupling picture of hydrodynamization (see Fig.~\ref{fig:IS_QM} and, e.g., \cite{Berges:2020fwq}). Current efforts aim to refine this picture further by extending studies to fully three-dimensional dynamics and embedding them in phenomenological frameworks. Second, how can these stages be probed experimentally, and what are their signatures? In this context, it is essential to study how the pre-QGP dynamics affects experimental observables. Particularly hard probes may provide new opportunities for small and large systems.

In these proceedings, we will summarize recent developments in the theoretical understanding of initial stages that were presented at Quark Matter 2025, discuss possible experimental signatures, and highlight new directions for future studies.

\section{Pre-equilibrium features of the initial stages}
\label{sec:features}

From a weak-coupling perspective $\lambda \ll 1$, the pre-QGP medium probes the very nature of quantum physics. As illustrated in Fig.~\ref{fig:IS_QM}, immediately after the collision, the system is dominated by strong gluon fields with high occupancy $\sim 1/\lambda$ that behave as classical waves and form the so-called Glasma \cite{Gelis:2012ri}. It is characterized by large gauge potentials and longitudinal flux tubes with transverse domains of size $\sim 1/Q_s$, with the saturation scale $Q_s$, that persist up to the time $\tau \lesssim 1/Q_s \sim 0.1 - 0.2$ fm/c. Later, the system evolves towards a QCD kinetic theory description in terms of scatterings of quasiparticles \cite{Arnold:2002zm} that describe the hydrodynamization process of the bottom-up scenario \cite{Baier:2000sb, Berges:2013eia, Kurkela:2015qoa}. The medium passes through dense and dilute stages and exhibits strong momentum anisotropies, with typical local thermalization timescales $\tau \lesssim 1$~fm/c. 
Although the Glasma and kinetic theory regimes are often matched, in reality, the transition is smooth passing by a phase of plasma instabilities \cite{Mrowczynski:1993qm, Romatschke:2005pm} before arriving at a shared regime where both descriptions are applicable \cite{Berges:2013eia, Berges:2013fga}. Importantly, the aforementioned distinct features of the Glasma and kinetic stages may, in principle, be experimentally testable. 
It should be emphasized that the picture presented here relies on weak-coupling assumptions and can be contrasted with the strong-coupling holographic approach~\cite{Berges:2020fwq}, since the more realistic coupling $\lambda \sim 10$ is neither weak nor strong.

\subsection{Initial stages go spatially 3D}

Recent progress has been made toward describing and understanding the spatial structure of the initial stages beyond boost-invariant approximations. Extensions of Glasma simulations to three dimensions reveal efficient descriptions of longitudinal structures in the dilute regime~\cite{Ipp:2021lwz, Ipp:2024ykh}, building on earlier results of the more general case in~\cite{Gelfand:2016yho, Ipp:2017lho}. Developments in kinetic theory include new approaches to solving the QCD evolution in full 3+3D (spatial and momentum) using machine learning techniques~\cite{BarreraCabodevila:2025ogv}. 
Hybrid frameworks are also advancing: the K{\o}MP{\o}ST approach that adds energy-momentum tensor fluctuations to a homogeneous QCD kinetic theory background \cite{Kurkela:2018wud} has been generalized to two and three spatial dimensions, in a multi-stage description of the evolution \cite{Garcia-Montero:2025bpn} complemented by the CGC-based McDipper for the initial state~\cite{Garcia-Montero:2023gex} and CLVisc for 3+1D viscous hydrodynamics. 
Moreover, Boltzmann-type transport approaches focusing on $2\leftrightarrow 2$ processes have been extended to 3+1D, showing the emergence of dynamical attractors in systems beyond the boost-invariant limit~\cite{Nugara:2024net}.

\subsection{Better understanding of the pre-QGP era}

At the level of kinetic theory, new insights have been obtained in simplified descriptions of the early stages. A reduced Boltzmann equation in the diffusion approximation for both elastic and inelastic scatterings reproduces the full QCD effective kinetic theory evolution during bottom-up hydrodynamization to reasonable accuracy, providing a more efficient and simple framework for longitudinally expanding systems~\cite{BarreraCabodevila:2025vir}. A deeper understanding of the hydrodynamization process can be gained by studying the eigenvalues and eigenmodes of an effective Hamiltonian of the kinetic description in the adiabatic hydrodynamization framework \cite{Brewer:2019oha}, which has now been extended to include inelastic scatterings~\cite{Rajagopal:2025nca}. 
Complementary studies \cite{DeLescluze:2025jqx} highlight the role of quasinormal modes that are related to those of the adiabatic hydrodynamization framework during the nonthermal fixed point \cite{Berges:2014bba} of the bottom-up stage. 
Finally, highly occupied plasmas have recently been shown to exhibit thermalization of magnetic modes triggered by chaotic dynamics~\cite{Pandey:2024goi}.

\section{Hard probes: window into the initial stages}
\label{sec:hard_probes}

Hard probes offer valuable opportunities to study the properties of the pre-equilibrium QGP. Electromagnetic probes, such as direct photons and dileptons, are produced throughout the evolution and escape the medium largely unmodified, making them sensitive to the earliest stages~\cite{Geurts:2022xmk}. In contrast, QCD probes including jets, heavy flavors, and quarkonia interact strongly with the medium at all times and therefore accumulate information from different stages of the evolution~\cite{Apolinario:2022vzg}.

{\em Electromagnetic probes} have received renewed attention as signatures of pre-equilibrium dynamics. Recent studies have shown that dileptons can provide information on the early stages \cite{Coquet:2021lca} and particularly on plasma anisotropies, with polarization emerging as a signal of early non-equilibrium distributions~\cite{Coquet:2023wjk}. 
Similarly, the polarization of photons has been suggested as a signature of the early stages \cite{Hauksson:2023dwh}.

\begin{figure}[t]
\centering
\includegraphics[height=3.5cm]{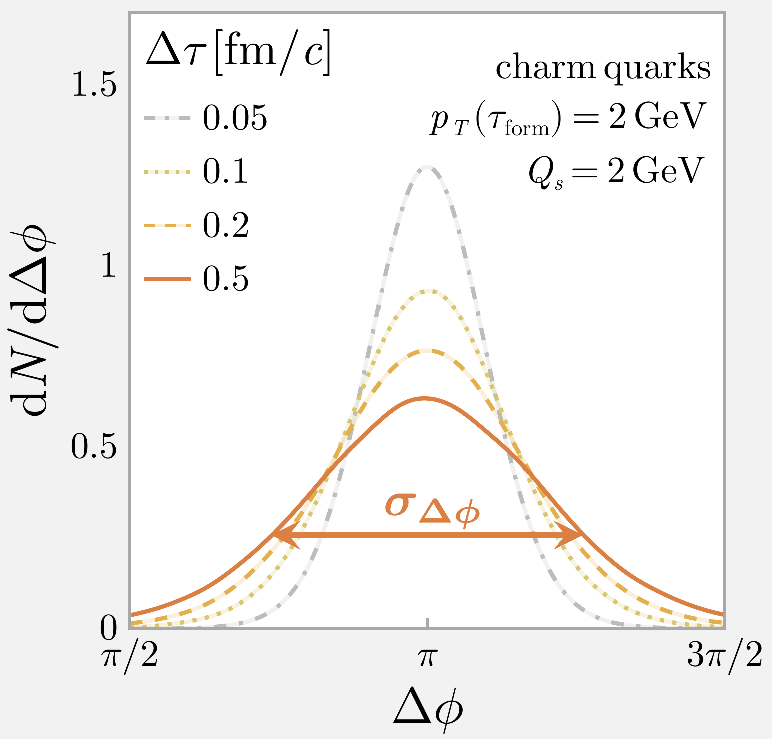}
\includegraphics[height=3.5cm]{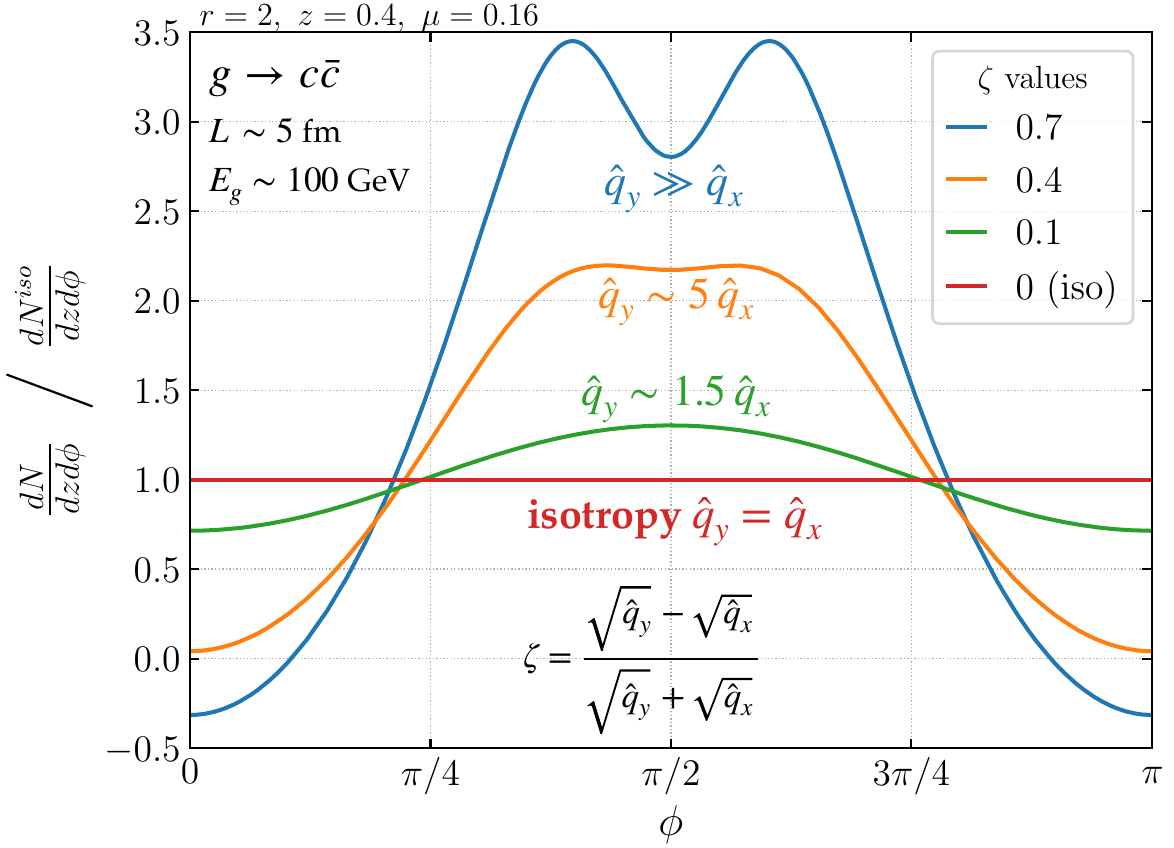}
\includegraphics[height=3.5cm]{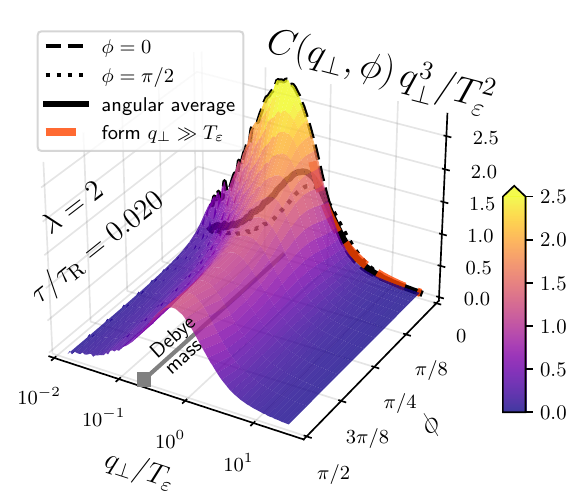}
\caption{{\em Left:} A sizable azimuthal correlation of $c \bar c$ pairs created in the Glasma remains even at relatively late times~\cite{Avramescu:2024xts, Avramescu:2024poa}. {\em Center:} One observes large deviations for the (differential) spectrum of a jet parton splitting into a $c \bar c$ pair in an anisotropic plasma as compared to the isotropic case~\cite{Barata:2024bqp}. {\em Right:} The jet collision kernel (that enters radiative energy loss and splitting rate calculations) of a pre-QGP plasma in the kinetic regime (using the isoHTL screening prescription \cite{Boguslavski:2024kbd}) shows strong azimuthal ($\phi$) and transverse momentum ($q_\perp$) variations at early times~\cite{Altenburger:2025iqa, Lindenbauer:2025ctw}.}
\label{fig:anisotropies}       
\end{figure}

\subsection{Jet and heavy-quark transport coefficients}

Once created, {\em jets and heavy quarks} interact with the medium and thus provide an important window into the properties of the non-equilibrium pre-QGP matter. Their evolution can be characterized by transport coefficients that quantify momentum broadening. For heavy quarks with masses $M \gg p$, the relevant quantity is the heavy-quark diffusion coefficient
$\kappa_i = \frac{\ud}{\ud \tau} \langle p_i^2 \rangle$, 
which describes stochastic momentum kicks and can enter Lindblad evolution equations for quarkonium dynamics. In analogy, the transverse momentum broadening of jets with $p \gg Q_s, T$ is described by the jet quenching parameter
$\hat q_i = \frac{\ud}{\ud \tau} \langle p_{\perp,i}^2 \rangle$ in the direction $i$. 
The time dependence of these coefficients during the pre-equilibrium evolution can leave observable imprints on hard probes: $\hat q_i(\tau)$ affects jet energy loss and substructure, while $\kappa_i(\tau)$ influences heavy-quark diffusion, nuclear modification factors, and the in-medium dynamics of quarkonia. 
Recent studies emphasize the importance of incorporating pre-equilibrium dynamics into phenomenological models, as transport coefficients from the early stages can impact the interpretation of jet quenching and heavy-flavor suppression in experimental data.

These transport coefficients have been computed in detail during the pre-equilibrium stages. In the Glasma phase, classical-statistical simulations allow the extraction of $\kappa_i$ and $\hat q_i$~\cite{Ipp:2020nfu, Ipp:2020mjc, Boguslavski:2020tqz, Carrington:2016mhd, Carrington:2021dvw, Carrington:2022bnv, Khowal:2021zoo, Avramescu:2023qvv, Pandey:2023dzz, Backfried:2024rub}, leading to large anisotropic values $\kappa_z > \kappa_T$ and $\hat q_z > \hat q_y$ with beam direction $z$.
A notable development is the identification of a mass dependence of $\kappa$, which extends spectral function studies from light \cite{Boguslavski:2021kdd} to heavy quarks~\cite{Pandey:2023dzz}. Spectral reconstruction methods, where gauge-invariant transport coefficients are compared with integrals over gauge-fixed spectral functions~\cite{Boguslavski:2020tqz, Backfried:2024rub}, have successfully explained the observed properties of their evolution. For example, an initial increase $\kappa_i \sim \tau$ is due to a coherence effect, while in two-dimensional Glasma-like systems the observed negative $\kappa_z < 0$ originates from narrow scalar excitations and the observed large values of $\kappa_T$ are additionally enhanced due to broad gluonic excitations and a novel transport peak \cite{Backfried:2024rub, Boguslavski:2021buh}.  
During the kinetic phase, $\hat q(\tau)$ seems to interpolate quite smoothly from Glasma to hydrodynamic values, whereas $\kappa$ exhibits much larger deviations~\cite{Boguslavski:2023fdm, Boguslavski:2024ezg, Boguslavski:2024jwr, Boguslavski:2023jvg}. Both coefficients retain the same anisotropic ordering as in the Glasma, with $\kappa_z > \kappa_T$ and $\hat q_z > \hat q_y$, however, each component connecting to the Glasma phase in a less smooth manner. Importantly, the contribution from the pre-equilibrium regime is not negligible: the total amount of heavy-quark momentum broadening is found to be of the same parametric order as in the hydrodynamic phase, $\langle \Delta p^2 \rangle_{\mathrm{neq}} \sim \langle \Delta p^2 \rangle_{\mathrm{eq}}$~\cite{Carrington:2021dvw, Boguslavski:2023fdm, Singh:2025duj}.

\subsection{Impact of QCD probes on observables}

The anisotropies and dynamics of the pre-QGP medium leave characteristic traces on observables. For instance, correlations of heavy-quark pairs in the Glasma, the spectrum of a jet parton splitting into a $c \bar c$ pair, and the jet-medium interaction potential during the kinetic regime all reveal nontrivial azimuthal dependences, as illustrated in Fig.~\ref{fig:anisotropies}~\cite{Avramescu:2024xts, Avramescu:2024poa, Barata:2024bqp, Boguslavski:2024kbd, Altenburger:2025iqa, Lindenbauer:2025ctw}. 
It was also suggested that medium anisotropies could manifest themselves in the polarization of jets~\cite{Hauksson:2023tze} or in the features of the differential gluon radiation rates~\cite{Barata:2025xxx}. Related studies have investigated medium-induced gluon radiation in Glasma models~\cite{Barata:2024xwy} and in evolving plasmas with time-dependent $\hat q(\tau)$~\cite{Adhya:2024nwx}.

Further developments have been made for heavy flavors and quarkonia. Predictions for $D$-meson observables have been obtained from IP-Glasma hybrid frameworks~\cite{Singh:2025duj}. Modified Wong equations have been used to describe quarkonium melting in the Glasma, including memory effects~\cite{Ruggieri:2022kxv, Pooja:2024rnn}. Pre-equilibrium effects have also been studied in smaller systems, such as charm and beauty diffusion in the Glasma in pA collisions, where considerable quarkonium melting has been seen already at the pre-equilibrium stage~\cite{Oliva:2024rex}.

\section{Conclusion}
\label{sec:concl}

The initial stages of heavy-ion collisions provide unique insights into the real-time dynamics of QCD. From a weak-coupling perspective, the pre-equilibrium medium evolves from coherent gluon fields to quasiparticles and kinetic theory. Recent work has extended this picture into full spatial dynamics and revealed novel features suitable for embedding into multi-stage frameworks. Hard probes---both electromagnetic and QCD---offer a promising way to test these developments, but require further studies on both theoretical and experimental frontiers. 
A recent theory workshop (\href{https://indico.cern.ch/event/1487879/}{``High energy probes of the initial stages''}) has additionally stimulated progress in this direction, which may provide compelling opportunities for the high-luminosity LHC Runs 4 and 5, small collision systems, and CBM-FAIR experiments.

\bibliography{refs}

\end{document}